\documentclass[twoside,11pt]{article}

\usepackage{jmlr2e}
\usepackage{tabto}
\usepackage{hyperref}
\usepackage{svg}
\usepackage{graphics}
\usepackage{graphicx}
\graphicspath{{}}
\usepackage{footmisc}
\usepackage[clock]{ifsym}
\usepackage{fourier}
\usepackage{fontawesome}

\usepackage{xcolor}
\usepackage{pifont}
\newcommand\na{{\ding{55}}}
\newcommand\valid{{\ding{51}}}
\newcommand\avalid{\textcolor{gray}{\ding{51}}}

\newcommand\ram{\faStackOverflow}

\ShortHeadings{Scikit-network: Graph analysis in Python}{Bonald, Charpentier, de Lara and Lutz}
\firstpageno{1}

\begin{document}

\title{Scikit-network: Graph Analysis in Python}

\author{\name Thomas Bonald \email thomas.bonald@telecom-paris.fr \\
        \name Nathan de Lara \email nathan.delara@telecom-paris.fr \\
        \name Quentin Lutz\thanks{\hspace{.2cm}Also affiliated with Nokia Bell  Labs, France} \email quentin.lutz@telecom-paris.fr \\
       \addr Institut Polytechnique de Paris\\
       France
       \AND
       \name Bertrand Charpentier \email bertrand.charpentier@in.tum.de\\
       \addr Technical University of Munich \\
       Germany
  }

\editor{Andreas Mueller}

\maketitle

\begin{abstract}%   <- trailing '%' for backward compatibility of .sty file
{\it Scikit-network} is a Python package inspired by 
scikit-learn for the analysis of large graphs. Graphs are represented by their adjacency  matrix in the sparse CSR format of SciPy. The package provides  state-of-the-art algorithms for ranking, clustering, classifying, embedding and visualizing the nodes of a graph. High performance is achieved through     a mix of  fast matrix-vector products (using SciPy),  compiled code (using Cython) and parallel processing. The package is distributed under the BSD license, with dependencies  limited to  NumPy and SciPy. It is compatible with Python 3.6 and newer. Source code, documentation and installation instructions are available online\footnote{  \url{https://scikit-network.readthedocs.io/en/latest/}}.
\end{abstract}


\section{Introduction}

Scikit-learn \citep{pedregosa2011scikit} is a machine learning package based on the popular Python language. It is well-established in today's machine learning community thanks to its versatility, performance and ease of use, making it suitable for  both researchers, data scientists and data engineers.  Its main assets are the variety of  algorithms,  the  performance of their implementation and their common API.

{\it Scikit-network} is a Python package inspired by scikit-learn for graph analysis. The sparse nature of real graphs, with up to millions of nodes, prevents their representation as dense matrices and rules out most algorithms of scikit-learn.  {\it Scikit-network}  takes as input a sparse matrix in the CSR format of SciPy and provides state-of-the-art algorithms  for ranking, clustering, classifying, embedding and visualizing the nodes of a graph. 
%The package is distributed under the BSD license.

The design objectives of {\it scikit-network}  are the same as those having made scikit-learn a success: versatility, performance and ease of use. The result is a  Python-native package, like NetworkX \citep{hagberg2008exploring}, that  achieves the state-of-the-art  performance 
of iGraph \citep{igraph} and graph-tool \citep{peixoto_graph-tool_2014}
(see the benchmark in section \ref{sec:perf}). 
{\it Scikit-network} uses the same API as Scikit-learn, with algorithms available as classes with the same methods (e.g., {\tt fit}). It is distributed with the BSD license, with    dependencies limited to  NumPy \citep{walt2011numpy} and SciPy \citep{2020SciPy-NMeth}.

\section{Software features}

\begin{figure}[h]
    \centering
    \def\svgwidth{\linewidth}
    \scalebox{0.47}{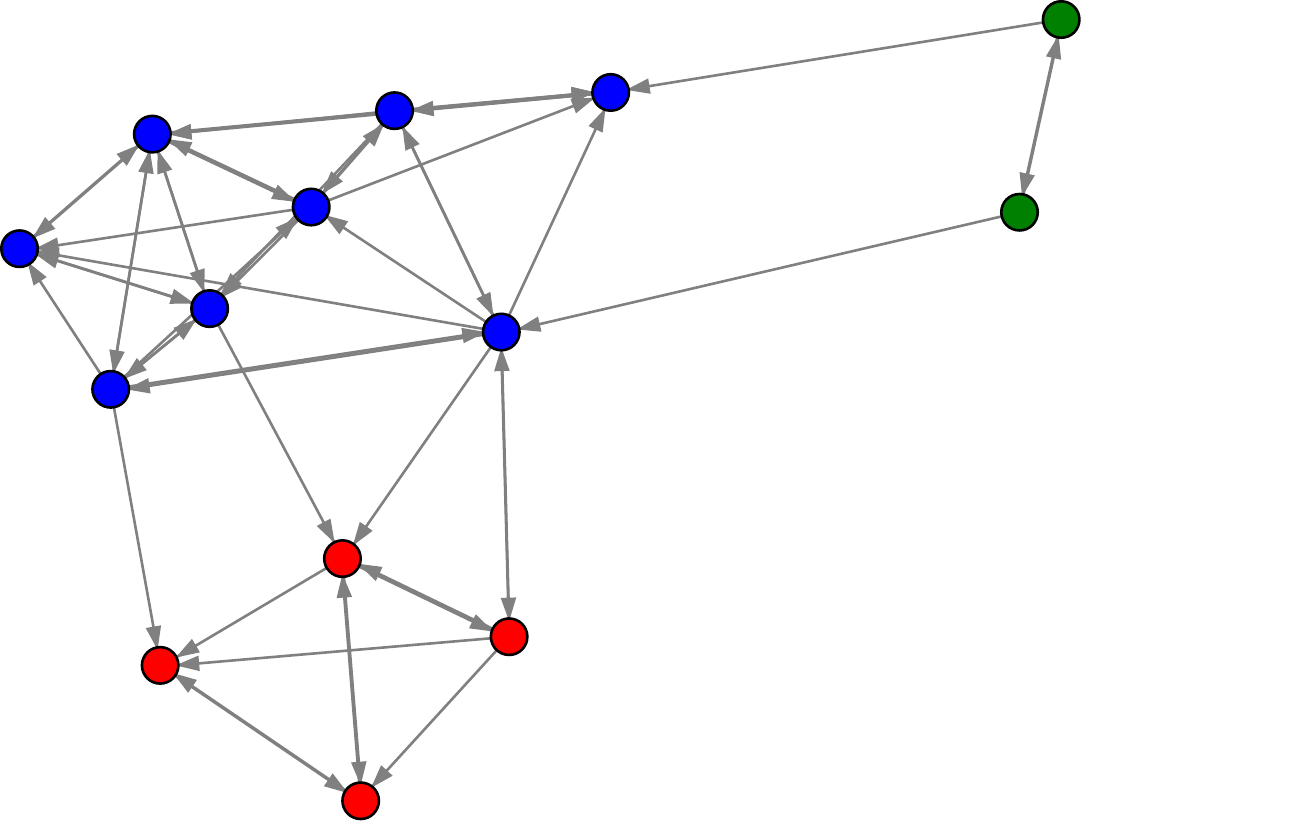}
    \scalebox{0.50}{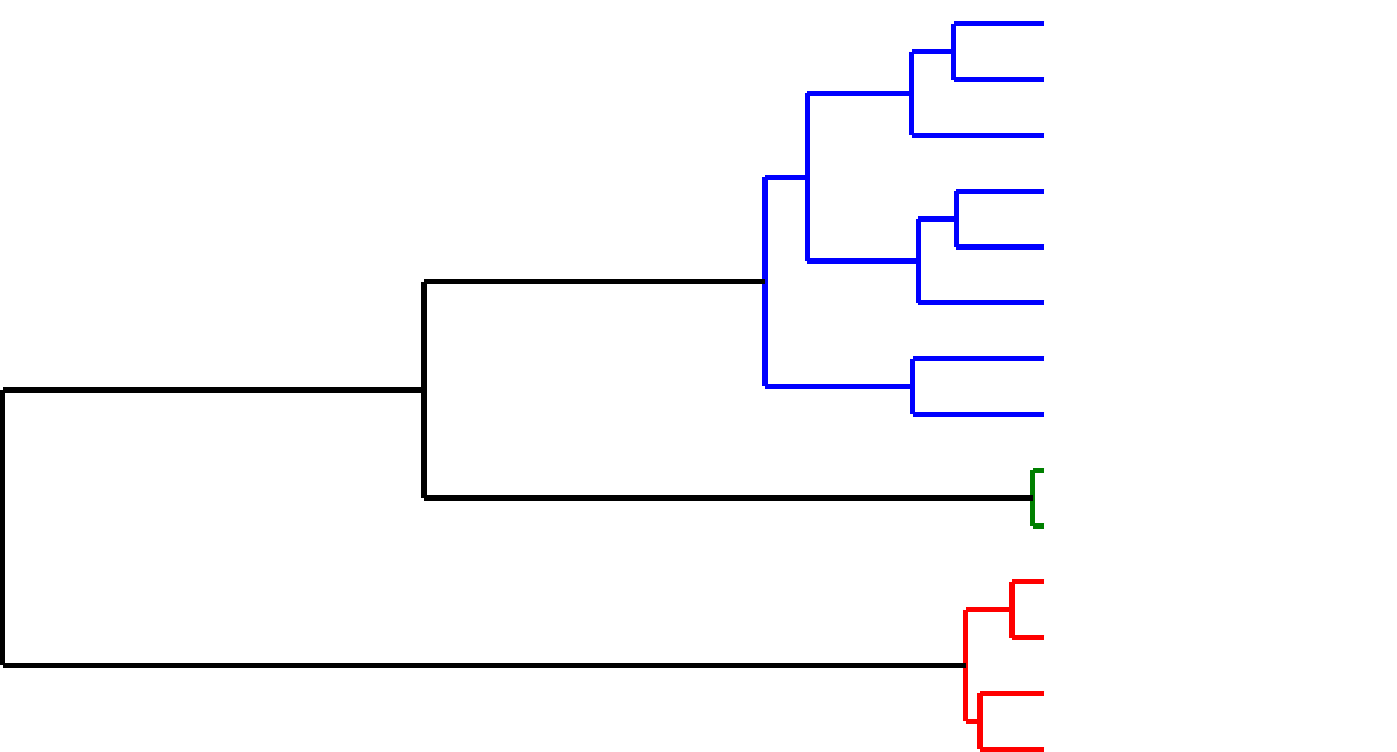}
    \caption{Visualization of a graph and a dendrogram  as SVG images.}
    \label{fig:visu}
\end{figure}

The package is organized in modules with consistent API, covering various  tasks:
\begin{itemize}

\item{\bf Data.} Module for loading graphs from distant repositories, including  Konect \citep{kunegis_konect:_2013}, parsing {\tt tsv} files into graphs, and generating graphs from standard models, like the stochastic block model \citep{airoldi2008mixed}.

\item{\bf Clustering.} Module for clustering graphs, including a  soft version that returns a node-cluster membership matrix. 

\item{\bf Hierarchy.} Module for the hierarchical clustering of graphs,  returning dendrograms in the standard format of  SciPy. The module also provides various post-processing algorithms for cutting and compressing dendrograms.

\item{\bf Embedding.} Module for embedding graphs in  a space of low dimension.  This includes spectral embedding and standard dimension reduction techniques like SVD and  GSVD, with key features like regularization. 

\item \textbf{Ranking.} Module for ranking the nodes of the graph by order of importance. This includes PageRank \citep{pagerank} and various centrality scores.

\item \textbf{Classification.}
Module for classifying the nodes of the graph based on the labels of a few nodes (semi-supervised learning). 

\item{\bf Connectivity.} Module relying on SciPy for the connectivity of the graph: shortest paths, graph traversals, connected components, etc. 

\item \textbf{Visualization.} Module for visualizing graphs and dendrograms in SVG (Scalable Vector Graphics) format.   Examples are displayed in Figure \ref{fig:visu}.
\end{itemize}

These modules are only partially covered by existing  graph softwares (see Table \ref{tab:feat}).  Another interesting feature of {\it scikit-network} is its ability to work directly on bipartite graphs, represented by their biadjacency matrix.

\begin{table}[h]
    \centering
    \begin{tabular}{l|cccc}
    Modules & scikit-network & NetworkX & iGraph & graph-tool\\
     \hline
        Data            & \valid{} & \avalid{}  & \na{} & \avalid{} \\
        Clustering      & \valid{} & \avalid{}  & \avalid{}  & \na{}\\
        Hierarchy       & \valid{} & \na{} & \avalid{}  & \avalid{}\\
        Embedding       & \valid{} & \avalid{}  & \na{} & \avalid{} \\
        Ranking         & \valid{} & \avalid{}   & \valid{}   & \valid{} \\
        Classification  & \valid{} & \avalid{}  & \na{} & \na{}\\
        Connectivity          & \valid{} & \valid{}   & \valid{}   & \valid{} \\
        Visualization   & \valid{} & \avalid{}  & \avalid{}  & \avalid{} \\
    \end{tabular}
    \caption{Overview of graph software features.
    \valid: Available. \avalid: Partially available or slow implementation.  \na: Not available.}
    \label{tab:feat}
\end{table}

\medbreak

\section{Project Assets}

\paragraph{\it Code quality and availability.}
Code quality is assessed by standard code coverage metrics. Today's coverage is at 98\% for the whole package. Requirements are also kept up to date thanks to the PyUp tool. {\it Scikit-network} relies on TravisCI for continuous integration and cibuildwheel and manylinux for deploying on common platforms. OSX, Windows 32 or 64-bit and most Linux distributions \citep{pep513} are supported  for Python versions 3.6 and newer.

\paragraph{\it Open-source  software.} The package is hosted on GitHub\footnote{\url{https://github.com/sknetwork-team/scikit-network}} and part of SciPy kits aimed at creating open-source scientific software. Its BSD license enables maximum interoperability with other software. Guidelines for contributing are described in the package's documentation\footnote{\label{foot:doc}\url{https://scikit-network.readthedocs.io/en/latest/}} and guidance is provided by the GitHub-hosted Wiki.

\paragraph{\it Documentation.}  {\it Scikit-network} is provided with a complete documentation\footref{foot:doc}. The API reference presents the syntax  while the tutorials present  applications on real graphs. Algorithms are documented with relevant formulas, specifications, examples and references, when relevant.

\paragraph{\it Code readability.} The source code follows the stringent PEP8 guidelines. Explicit variable naming and type hints make the code easy to read. The number of  object types is  kept to a minimum.

\paragraph{\it Data collection.} The package offers multiple ways to fetch data. Some small graphs are embedded in the package itself for testing or teaching purposes. Part of the API makes it possible to fetch data from selected graph databases easily. Parsers are also present to enable users to import their own data and save it in a convenient format for later reuse.

\section{Resources}
{\it Scikit-network} relies on a very limited number of external dependencies for ease of installation and  maintenance. Only SciPy and NumPy are required on the user side.

\paragraph{\it SciPy.} Many elements from SciPy are used for both high performance and simple code. The sparse matrix representations allow for efficient manipulations of large graphs while the linear algebra solvers are used in many algorithms. Scikit-network also relies on the \textit{LinearOperator} class for efficient implementation of certain algorithms.

\paragraph{\it NumPy.} NumPy arrays are used through SciPy's sparse matrices for memory-efficient computations. NumPy is used throughout the package for the manipulation of  arrays. Some inputs and most of the outputs are given in the NumPy array format.

\paragraph{\it Cython.}
In order to speed up execution times, Cython     \citep{behnel2011cython} generates C++ files automatically using a Python-like syntax. Thanks to the Python wheel system, no compilation is required from the user on most platforms. Note that Cython has a built-in module for parallel computing on which {\it scikit-network} relies for some algorithms. Otherwise, it uses Python's native multiprocessing.

\section{Performance}
\label{sec:perf}

To show the performance of {\it scikit-network}, we compare the implementation
of some representative algorithms with those of the graph softwares of Table \ref{tab:feat}:
the Louvain clustering algorithm \citep{blondel_fast_2008}, PageRank \citep{pagerank}, HITS \citep{kleinberg1999authoritative} and the spectral embedding \citep{belkin2003laplacian}.
For PageRank, the number of iterations is set to $100$ when possible (that is for all packages except iGraph).
For Spectral, the  dimension of the embedding space is set to $16$.

Table \ref{tab:benchmark} gives the running times of these algorithms on 
the \textit{Orkut} graph of  Konect \citep{kunegis_konect:_2013}.
The graph has $3,072,441$ nodes and $117,184,899$ edges. The computer has a Debian 10 OS and is equipped with an AMD Ryzen Threadripper 1950X 16-Core Processor and 32 GB of RAM. %The maximum authorized running time is set to one hour.  
As we can see, {\it scikit-network} is highly competitive.

\begin{table}[ht]
    \centering
    \begin{tabular}{l|cccc}
         & scikit-network & NetworkX & iGraph & graph-tool\\
         \hline
        Louvain     & 771 & \na   & 1,978   & \na      \\
        PageRank    & 48 & \ram  & 236   & 45      \\
        HITS        & 109 & \ram  & 80     & 144     \\
        Spectral    & 534 & \ram  & \na     & \na    
    \end{tabular}
    \caption{Execution times (in seconds). \na: Not available. \ram: Memory overflow. }
    \label{tab:benchmark}
\end{table}

We also give in Table \ref{tab:memory}  the memory usage of each package when loading the graph. Thanks to the  CSR format, {\it scikit-network} has a minimal footprint.

\begin{table}[ht]
    \centering
    \begin{tabular}{l|cccc}
         & scikit-network & NetworkX & iGraph & graph-tool\\
         \hline
        RAM usage     & 1,222 %1\,251 
        &  \ram  & 17,765 %18\,191   
        & 10,366 %10\,615    
    \end{tabular}
    \caption{Memory usage (in MB). \ram: Memory overflow.}
    \label{tab:memory}
\end{table}

\acks{This work is supported by Nokia Bell Labs (CIFRE convention 2018/1648).}


\vskip 0.2in
\bibliography{sample}

\end{document}